\documentclass{aa}  
\usepackage{graphicx}
\usepackage{txfonts}
\begin{document} 
   \title{Dissipationless collapse and the dynamical mass-ellipticity relation of elliptical galaxies in Newtonian gravity and MOND}
\titlerunning{Dissipationless collapse and the dynamical mass ellipticity relation}
   \author{Pierfrancesco Di Cintio
          \inst{1,2,3,4}
         }          
   \institute{National Council of Research - Institute of Complex Systems (CNR-ISC), Via Madonna del piano 10, I-50019 Sesto Fiorentino, Italy
         \and
             National Institute of Nuclear Physics (INFN) -  Florence unit, via G.\ Sansone 1, I-50019 Sesto Fiorentino, Italy
          \and 
             Department of Physics and Astronomy, University of Florence, Piazzale E.\ Fermi 2, I-50125 Firenze, Italy
           \and
           National Institute of Astrophysics - Arcetri Astrophysical Observatory (INAF-OAA), Piazzale E.\ Fermi 5, I-50125 Firenze, Italy\\
           \email{pierfrancesco.dicintio@cnr.it}
             }

   \date{Received October 12, 2023; accepted February 5, 2024}

 
  \abstract
   {Deur (2014) and Winters et al. (2023) proposed an empirical relation between the dark to total mass ratio and ellipticity in elliptical galaxies from their observed total dynamical mass-to-light ratio data $M/L = (14.1 \pm 5.4)\epsilon$. In other words, the larger is the content of dark matter in the galaxy, the more the stellar component would be flattened. Such observational claim, if true, appears to be in stark contrast with the common intuition of the formation of galaxies inside dark halos with reasonably spherical symmetry.}
   {Comparing the processes of dissipationless galaxy formation in different theories of gravity, and emergence of the galaxy scaling relations therein is an important frame where, in principle one could discriminate them.}
   {By means of collisionless $N$-body simulations in modified Newtonian dynamics (MOND) and Newtonian gravity with and without active dark matter halos, with both spherical and clumpy initial structure, I study the trends of intrinsic and projected ellipticities, S\'ersic index and anisotropy with the total dynamical to stellar mass ratio.}
   {It is shown that, the end products of both cold spherical collapses and mergers of smaller clumps depart more and more from the spherical symmetry for increasing values of the total dynamical mass to stellar mass, at least in a range of halo masses. The equivalent Newtonian systems of the end products of MOND collapses show a similar behaviour. The $M/L$ relation obtained from the numerical experiments in both gravities is however rather different from that reported by Deur and coauthors.}
   {}


   \maketitle
%

\section{Introduction}
In the $\Lambda$CDM scenario, stellar systems such as galaxies and clusters are thought to be embedded in Dark Matter (hereafter DM) halos, accounting for the missing mass fraction evidenced by velocity dispersion or rotational velocity measures (e.g. see \citealt{2014dyga.book.....B}). DM halos are usually assumed to be spherically symmetric structures in collisionless equilibrium with radial densities well fitted by the \cite{Navarro_etal_1997} (hereafter NFW) profile, originally obtained from cosmological dissipationless collapse simulations (see  \citealt{1991ApJ...378..496D}).\\
\indent In the context of spheroids such as elliptical galaxies and the bulges of disk galaxies the interplay between the DM density distribution $\rho_{\rm DM}$ and that of the of the stellar component $\rho_*$ has attracted a lot of interest, in particular with respect to the widely studied cusp-core problem (\citealt{1994Natur.370..629M}). That is, the observed rotation curves of dwarf galaxies imply a flat-cored halo (i.e. vanishing logarithmic density slope at inner radii) in contrast with the cuspy (i.e. diverging DM density at $r\rightarrow 0$) distributions predicted by DM $N-$body simulations, such as for example the NFW profile (e.g. see \citealt{Oh_2015}). In addition, it can also be proved analytically that flat cored models can not admit consistent phase-space distributions when embedded in cuspy halos, except in a small range of values of the central anisotropy profile (see \citealt{1999ApJ...520..574C}, see also \citealt{1992MNRAS.255..561C}).\\
\indent Several explanations to the core-cusp problem have been proposed so far, from the effect of stellar evolution feedback (\citealt{2014MNRAS.437..415D,2014ApJ...784L..14B,2018MNRAS.474.1398G,2018MNRAS.480.4287K,2021ApJ...921..126B}) to extra channels for dissipation, implying that the flattening of the central DM density distribution, is due the DM's intrinsic nature (e.g. the so-called  self-interacting DM, \citealt{2012JCAP...02..014B,2019JCAP...12..010K,2021MNRAS.504.2832S,2022MNRAS.513.3458B}). Some authors also suggested a misinterpretation of the observational data on dwarf galaxies (\citealt{2003ApJ...584..566M,2007ApJ...657..773V}). Moreover, in the context of modified theories of gravity, as alternatives to the DM hypothesis, such as for example the modified Newtonian dynamics (hereafter MOND, \citealt{1983ApJ...270..365M}) explain the core-cusp problem as an artifact of the gravity model (see e.g. \citealt{2021A&A...656A.123E,2022ApJ...940...46S,2023arXiv230708865R}).\\
\indent If on one hand much work has been devoted to the interplay between the slopes of the dark and visible matter density profiles and their implications for the anisotropy profile, on the other much less is known on the relationship of the DM profile and the intrinsic shape of the stellar component, namely its ellipticity $\epsilon$. \cite{2014MNRAS.438.1535D,2020arXiv201006692D} and more recently \cite{2023MNRAS.518.2845W}, using a broad sample of elliptical galaxies from independent surveys, and different methods to evaluate the mass to light ratio $M/L$ (i.e. Virial theorem \citealt{1985A&A...152..315B}; anisotropic Jeans modelling \citealt{2008MNRAS.390...71C,2017ApJ...850...15P}; gravitational lensing \citealt{2008ApJ...682..964B,2010ApJ...724..511A,2017ApJ...851...48S}; gas disk dynamics \citealt{1991ApJ...373..369B,1993ApJ...416L..45B,1997A&A...323..349P}; X-ray emission spectra \citealt{2020MNRAS.491.1690J} and the dynamics of satellite star clusters or companion galaxies \citealt{2017MNRAS.468.3949A,2019ApJ...887..259H,2020ApJ...903...38C}), observed that the two quantities are related by the linear relation
\begin{equation}\label{epstoml}
  M/L = (14.1 \pm 5.4)\epsilon. 
\end{equation}
In the equation above, the mass to light ratio is normalized such that $M/L(\epsilon_{\rm 2D}=0.3)\equiv 8M_{\odot}/L_{\odot}\equiv 4M/M_*(\epsilon_{\rm 2D}=0.3)$, and the intrinsic ellipticity $\epsilon$ is inferred from its apparent 2D-projected value $\epsilon_{\rm 2D}$ in the assumption of oblateness and a Gaussian distribution of projection angles $\theta$ from
\begin{equation}\label{epsilonapp}
\epsilon_{\rm 2D}=1-\sqrt{(1-\epsilon)^2\sin^2\theta+\cos^2\theta}.
\end{equation}
Equation (\ref{epstoml}) implies that (at least for the galaxies examined) a larger contribution of the DM mass $M_{\rm DM}$ to the total mass $M$ corresponds to a larger departure from the spherical symmetry (i.e. larger ellipticity). The lower DM fraction in rounder galaxies could be, in principle, perceived as being in stark contrast with the standard scenario of galaxy formation\footnote{Note that, on the contrary, globular clusters are essentially spherical while not containing a significant fraction of DM}, requiring that the baryons aggregates and forms structure in dark matter halos that had collapsed at earlier times. Notably, however, some peculiar elliptical galaxies (though excluded by the original sample of \citealt{2023MNRAS.518.2845W}) such as the dwarf (\citealt{1998ARA&A..36..435M}) or ultrafaint dwarf galaxies (\citealt{2019ARA&A..57..375S}) appear to go against the trend given by Eq. (\ref{epstoml}) as they are characterized by a rather spherical shape ($\epsilon\leq 0.1$) while being DM dominated with $M/L$ up to $10^2$ or more. Of course, one of the reasons behind the mass ratio - ellipticity relation could be related to the departure from spherical symmetry of the halos themselves and its relation to their total DM content. In fact, several studies (both observational and numerical, see e.g. \citealt{2002ApJ...574..538J,2010MNRAS.407..581R,2011MNRAS.416.1377V,2012MNRAS.420.3303B,2012JCAP...05..030S,Rojas-Niño_2015,2016ApJ...826L..23E,2022MNRAS.517.4827G}) point out that more massive halos have significantly higher departure from the spherical symmetry. More recently, \cite{2023arXiv230708865R} performed $N-$body simulations in MOND showing that the equivalent Newtonian systems (ENS, i.e. stellar systems with a dark halo such that the total Newtonian gravitational potential is the same as the parent MOND model) of the end products of cold gravitational collapses are consistent with Eq. (\ref{epstoml}) if the initial conditions are sufficiently clumpy.\\
\indent The origin of the triaxiality of (single component) gravitational systems, forming via direct collapse from nearly spherical initial conditions, is usually ascribed to the process of radial-orbit instability\footnote{Some authors, (see \citealt{ejection_mjbmfsl,2015MNRAS.446.1335W,2015MNRAS.449.4458S,2016A&A...585A.139B}) also suggest the angular momentum loss via particle escape during the violent phases of gravitational collapse as a viable channel for the loss of spherical symmetry.} (hereafter ROI, see \citealt{Polyachenko_Shukhman_1981,1987MNRAS.224.1043P,1994ApJ...434...94B,2010MNRAS.405.2785M}), where an initially minor departure from spherical symmetry grows by accreting particles on low angular momentum orbits (\citealt{1990MNRAS.242..576A}). This finds confirmation in numerical experiments; \cite{merritt+aguilar_1985,2008ApJ...685..739B,2009ApJ...704..372B,2015MNRAS.447...97G,2017MNRAS.468.2222D,2020MNRAS.494.1027D,2022arXiv220906846W}.\\
\indent The ROI is stronger in models with a larger degree of radial anisotropy\footnote{ Though extensively studied for anisotropic equilibrium models (\citealt{2011MNRAS.416.1836P} and references therein), the ROI is also at play in initially far-from-equilibrium setting (\citealt{2005A&A...433...57T}) 
 undergoing violent relaxation (\citealt{lyndenbell}), such as for example the early phases of the collapse of initially cold systems.}, quantified (see e.g. \citealt{2008gady.book.....B}, see also \citealt{1984sv...bookR....F}) by the parameter 
\begin{equation}
\xi=\frac{2K_r}{K_t},
\end{equation}
where $K_r$ and $K_t=K_\theta+K_\phi$ are the radial and tangential components of the kinetic energy tensor, respectively given by
\begin{equation}
K_r=2\pi\int\rho(r)\sigma^2_r(r)r^2{\rm d}r,\quad K_t=2\pi\int\rho(r)\sigma^2_t(r)r^2{\rm d}r,
\end{equation} 
where $\sigma^2_r$ and $\sigma^2_t$ are the radial and tangential phase-space averaged square velocity components. Typically, single component models have been fount to be (numerically) stable for $\xi\lesssim 1.7$ (e.g. see \citealt{2002MNRAS.332..901N}) while some authors put a larger threshold for stability at $\xi\lesssim 2.3$ (see \citealt{1997ApJ...490..136M}, see also \citealt{2011TTSP...40..425M}).\\
\indent In stellar systems with a DM halo, the latter has been often indicated as having a stabilizing effect against ROI, however numerical experiments seem to weaken this hypothesis (\citealt{1991ApJ...382..466S,2011MNRAS.414.3298N}), while in general, even if the total amount of anisotropy is large, an extended isotropic core does indeed reduce the ROI, as shown by \cite{2006ApJ...637..717T}. Moreover, a locally fluctuating halo potential has in principle different effects on the ignition of the ROI at fixed $\xi$ for different baryon profiles, as conjectured by \cite{proceedingROI}.\\
\indent Simple $N-$body experiments of dissipationless collapse, merging, or instability growth in anisotropic equilibria have been rather successful in reproducing the scaling laws of elliptical galaxies and the fundamental plane itself (see for example the work by \citealt{2002MNRAS.332..901N,2003MNRAS.342..501N,2006MNRAS.370..681N}). Thus, it is natural to ask whether highly idealized collisionless numerical simulations could reproduce Eq. (\ref{epstoml}). In this work, I investigate by means of collisionless $N-$body simulations in Newtonian and MOND gravities the implications of the dissipationless collapse in the emergence of the $M/L$ - $\epsilon$ relation. The paper is structured as follows. In Section 2 I discuss the initial conditions, the numerical codes and the analysis of the simulation outputs. In Section 3 the results of the simulations are presented and interpreted. Section 4 draws the conclusions. 
\section{Models and methods}
\subsection{Properties of initial conditions}
The numerical simulations of dissipationless collapse presented in this paper (in both Newtonian and MOND paradigms) are characterized by two types of initial conditions, namely spherically symmetric initial states or clumpy baryon distributions. In all Newtonian simulations, when present, the DM halos are always initially spherical. Such halos have been implemented either with particles or with a non evolving smooth density distribution exerting a fixed potential. All simulated systems are assumed to be isolated so that any effect of the environment in their shape can be excluded, and in accordance with the explicit exclusion of group galaxies in the samples analyzed by \cite{2014MNRAS.438.1535D} and \cite{2023MNRAS.518.2845W}.\\
\indent The positions for the particles of the given component (baryons or DM) in spherical systems were sampled from the so called $\gamma-$models (\citealt{1993MNRAS.265..250D}, see also \citealt{1994AJ....107..634T}) defined by the density profile
\begin{equation}\label{dehnen}
\rho_i(r)=\frac{3-\gamma}{4\pi}\frac{M_{i}r_{c,i}}{r^\gamma(r+r_{c,i})^{4-\gamma}}.
\end{equation}
The density $\rho_i$ above is therefore defined by the total mass $M_i$, scale radius $r_{c,i}$ and logarithmic density slope $\gamma$. In this work I discuss mainly the $\gamma=0$ and $\gamma=1$ (i.e. the \citealt{1990ApJ...356..359H} model) cases, corresponding to a flat-cored and a moderately cuspy distribution, respectively. Using the density distribution (\ref{dehnen}), instead of the widely adopted NFW (or its alternatives, such as for example the \citealt{1965TrAlm...5...87E} profile, that can model a finite cusp or core, see also \citealt{2012A&A...540A..70R}), keeps the number model parameters limited to the density slope $\gamma$ one the mass ratio $M/M_*$ and scale radii are fixed, while preserving both the ''cosmological" central density $\rho_0(r)\propto 1/r$ and the large radii $1/r^3$ fall off.\\
\indent Similarly to previous studies (e.g. see \citealt{2003MSAIS...1...18L,2006MNRAS.370..681N,2013MNRAS.431.3177D} and references therein) baryonic particle velocities were sampled from a position independent isotropic Maxwell-Boltzmann distribution and then normalized in order to get the wanted value of the virial ratio 
\begin{table*}
\caption{Summary of the simulation properties: after the Gravitation theory (Newtonian or MOND Col. 1), we report the  number of particles representing baryons (Col. 2) and Dark Matter (Col. 3), the mass ratio (Col. 4), the initial profile or baryons (Col. 5) the Dark matter  (Cols. 6)  and the ratio of their scale radii the S\'ersic (Col. 7).} 
\begin{tabular}{llllllll}
\hline
&(1)  & (2)  & (3)  & (4)  & (5)  & (6)  & (7) \\
& Gravity &  $N_*$ &  $N_{DM}$ &  $M/M_*$ &  $\rho_*$ &  $\rho_{\rm DM}$ &  $r_{c*}/r_{c,DM}$\\
\hline 
(1) & Newton & $1.2\times 10^5$ &  /  &  1 & $\gamma=1$ & No DM  &  /  \\
(2) & Newton & $5\times 10^5$ &  /  &  1 & $\gamma=1$ & No DM  &  /  \\
(3) & Newton & $5\times 10^5$ &  /  &  1 & Clumpy & No DM  &  /  \\
(4) & Newton & $10^4$ &  /  &  1 & $\gamma=1$ & No DM  &  /  \\
(5) & Newton & $10^4$ &  /  &  1 & $\gamma=0$ & No DM  &  /  \\
(6) & Newton & $10^4$ &  /  &  1 & Clumpy & No DM  &  /  \\
(7) & Newton & $10^4$ &  Frozen halo  &  2, 4, 10, 30, 100 & $\gamma=1$ & $\gamma=1$  &  1  \\
(8) & Newton & $10^4$ &  Frozen halo  &  2, 4, 10, 30, 100 & $\gamma=1$ & $\gamma=1$  &  10  \\
(9) & Newton & $10^4$ &  Frozen halo  &  2, 4, 10, 30, 100 & $\gamma=0$ & $\gamma=0$  &  1  \\
(10) & Newton & $10^4$ &  Frozen halo  &  2, 4, 10, 30, 100 & Clumpy & $\gamma=1$  &  1, 10  \\
(11) & Newton & $10^4$ &  Frozen halo  &  50 & $\gamma=1$ & $\gamma=0$, 0.5, 1, 1.5  &  1  \\
(12) & Newton & $10^4$ &  Frozen halo  &  50 & $\gamma=1$ & $\gamma=0$, 0.5, 1, 1.5  &  10  \\
(13) & Newton & $1.2\times10^5$ &  Frozen halo  &  2, 4, 6, 8, 10, 50, 70 & $\gamma=1$ & $\gamma=1$  &  1  \\
(14) & Newton & $5\times 10^5$ &  Frozen halo  &  2, 3, 4, 5, 6, 7, 8, 9, 10, 11, 12, 13, 14, 15, 16, 17 & $\gamma=1$ & $\gamma=1$  &  1  \\
& &  &    &  18, 19, 20, 21, 22, 23, 24, 25, 26, 27, 28, 29, 30 &  &   &   \\
(15) & Newton & $5\times 10^5$ &  Frozen halo  &  2, 3, 4, 5, 6, 7, 8, 9, 10, 11, 12, 13, 14, 15, 16, 17 & Clumpy & $\gamma=1$  &  1  \\
& &  &    &  18, 19, 20, 21, 22, 23, 24, 25, 26, 27, 28, 29, 30 &  &   &   \\
(16) & Newton & $5\times 10^5$ &  $5\times 10^5$  &  2, 3, 4, 5, 6, 7, 8, 9, 10, 11, 12, 13, 14, 15, 16, 17 & $\gamma=1$ & $\gamma=1$  &  1  \\
& &  &    &  18, 19, 20, 21, 22, 23, 24, 25, 26, 27, 28, 29, 30 &  &   &   \\
(17) & Newton & $5\times 10^5$ &  $5\times 10^5$  &  2, 3, 4, 5, 6, 7, 8, 9, 10, 11, 12, 13, 14, 15, 16, 17 & Clumpy & $\gamma=1$  &  1  \\
& &  &    &  18, 19, 20, 21, 22, 23, 24, 25, 26, 27, 28, 29, 30 &  &   &   \\
(18) & Newton & $10^4$ &  $10^4$  &  2, 4, 10, 30, 100 & $\gamma=1$ & $\gamma=1$  &  1  \\
(19) & Newton & $10^4$ &  $10^4$  &  2, 4, 10, 30, 100 & Clumpy & $\gamma=1$  &  1  \\
(20) & Newton & $10^4$ &  $1,2,5\times 10^4$   &  5 & $\gamma=1$ & $\gamma=1$  &  1  \\
       &        &  $1,3\times 10^5$   &    &            &             &     \\
(21) & Newton & $10^4$ &  $1,2,5\times 10^4$   &  5 & $\gamma=0$ & $\gamma=0$  &  1  \\
       &        &  $1,3\times 10^5$   &    &            &             &     \\
(22) & MOND   & $5\times 10^5$ &  /  &  1.7, 2.3, 2.7, 3.5, 7.5, 14, 20, 25, 33, 47, 67, 81 & $\gamma=1$ & No DM  &  /  \\
&       &                &     &  107, 128, 153                                       &            &    &     \\
(23) & MOND   & $5\times 10^5$ &  /  &  1.7, 2.3, 2.7, 3.5, 7.5, 14, 20, 25, 33, 47, 67, 81 & Clumpy & No DM  &  /  \\
&       &                &     &  107, 128, 153                                       &            &    &     \\
\hline
\end{tabular}
\label{tab_sim}
\end{table*}
in the range $0\leq 2K/|W|\leq 0.2$. The simulations shown here are limited to the $2K/|W|=10^{-3}$. When considering a virialized live halo with (initial) density $\rho_{DM}$, the DM particles' velocities are obtained sampling with the usual rejection method the ergodic phase-space distribution function $f(\mathcal{E})$ evaluated numerically with the standard \cite{1916MNRAS..76..572E} inversion as
\begin{equation}\label{eddi}
f(\mathcal{E})=\frac{1}{\sqrt{8}\pi^2}\int_\mathcal{E}^{0}\frac{{\rm d}^2\rho_{DM}}{{\rm d}\Phi^2}\frac{{\rm d}\Phi}{\sqrt{\Phi-\mathcal{E}}},
\end{equation}
where, $\mathcal{E}=v^2/2-\Phi(r)$ is the specific energy for unit mass and the total potential  $\Phi=\Phi_*+\Phi_{\rm DM}$, where the stellar and DM potentials $\Phi_*$ and $\Phi_{\rm DM}$ are given for a $\gamma$-model by 
\begin{equation}\label{potgamma}
    \Phi_i(r)=\frac{GM_{i}}{r_{c,{i}}(2-\gamma)}\left[\left(\frac{r}{r+r_{c,{i}}}\right)^{2-\gamma}-1\right];\quad i=*,{\rm DM}.
\end{equation}
Clumpy initial conditions have been implemented following \cite{hansen_etal}. First a root $\gamma=1$ model is generated and then $N_C$ clumps also described by the density (\ref{dehnen}) with different choices of the scale radius $r_{c}$, and density slope $\gamma$ are generated with centres having Poissonian displacements from the root Hernquist model. When present, the DM halo is initialized as above for the spherical collapse. Again, stellar particles velocities are selected from a isotropic Maxwellian and later normalized to obtain the desired value of $2K/|W|$. MOND systems, as they lack of a dark component, are characterized by the dimensionless $\kappa$ parameter, defined by $\kappa\equiv GM/a_0r_c^2$, where $a_0\approx 10^{-8}$cm s$^{-2}$ is the MOND scale acceleration (\citealt{1983ApJ...270..365M}), so that for $\kappa\gg 1$ one recovers the Newtonian regime, while for $\kappa\lesssim 1$ the system is in MONDian regime.\\
\indent Throughout this work a constant mass to light ratio for the stellar component is adopted so that, in computer units $M_*/L=1$ for all models. As a consequence, Equation (\ref{epstoml}) is hereafter expressed in terms of $M/M_*$. All properties of the initial conditions presented in this work are summarized in Tab. \ref{tab_sim}.
\subsection{Numerical codes}
The Newtonian $N-$body simulations discussed here have been performed with the publicly available {\sc fvfps} code ({\sc fortran} version of a fast Poisson
solver, \citealt{2003MSAIS...1...18L}), using a parallel version of the classical \cite{1986Natur.324..446B} tree scheme for the force evaluation combined with the \cite{2002JCoPh.179...27D} fast multipole method (see also \citealt{dehnen+reed_2011}). The simulations span a range of $N=N_*+N_{\rm DM}$ (i.e. the total number of simulation particles) between $10^4$ and $10^6$ (cfr. Cols 2 and 3 in Tab. \ref{tab_sim} below). In the lowest resolution cases (here $N=10^4$) only stars are simulated with particles and the DM halo (when present) is always modelled as an external fixed potential (cfr. Eq. \ref{potgamma}).\\
\indent The gravitational forces are smoothed below a cut-off distance given by the so-called softening length (see \citealt{dehnen_2001}) in the range $0.02\leq\epsilon_{\rm soft}\leq 0.05$ in units of the initial $r_{c,*}$.\\
\indent MOND systems have been simulated using the {\sc nmody} modified Poisson Solver (for the details see \citealt{2011ascl.soft02001L}, see also \citealt{2007ApJ...660..256N}) applying an iterative relaxation procedure on a spherical grid\footnote{As a rule, here the grid was fixed to $128\times 32\times 64$ nodes.}, starting from a seed guess solution (in the same fashion of the standard Newtonian Poisson solvers see \citealt{1990MNRAS.242..595L,1991MNRAS.250...54L}) to evaluate the gravitational potential from the non-linear Poisson equation (\citealt{1984ApJ...286....7B})
\begin{equation}
\label{MOND}
   \nabla\cdot\left[\mu_M\left(\frac{||\nabla\Phi||}{a_0}\right)\nabla\Phi\right]=4\pi G\rho.
\end{equation}%
The MOND interpolation function $\mu(x)$ is assumed hereafter to be
\begin{equation}
\label{mu choices}
\mu_M(x)=\frac{x}{\sqrt{1+x^2}},
\end{equation}
yielding the usual asymptotic limits
\begin{equation}
\label{limits}
\mu_M(x)\sim
\begin{cases}
\displaystyle 1,\quad x\gg 1,\\
\displaystyle x,\quad x\ll 1;
\end{cases}
\end{equation}
that for $||\nabla\Phi||\gg a_0$ Eq. (\ref{MOND}) essentially recovers the Poisson Equation of the Newtonian regime, while for $||\nabla\Phi||\ll a_0$ the system is in the so-called deep-MOND (often labelled as dMOND) regime\footnote{It is worth noticing that it is not guaranteed that a given system is at all radii in dMOND regime, nor that when evolving from an initial condition in such state, it will continue to be so, \citealt{2007ApJ...660..256N}.} with Eq. (\ref{MOND}) simplifying to 
\begin{equation}
\label{dMOND}
   \nabla\cdot\left[||\nabla\Phi||\nabla\Phi\right]=4\pi G\rho a_0.
\end{equation}
\indent All simulations were extended up to $t=300t_{\rm Dyn}$ where as usual we define the half mass dynamical time 
\begin{equation}\label{tdyn}
  t_{\rm Dyn}\equiv\sqrt{2r_{h}^3/GM},  
\end{equation} 
for which $r_h$ is the radius containing half of the total system mass $M=M_*+M_{DM}$ in Newtonian models or $M=M_*$ in MOND models and single component Newtonian models. By doing so, one is ensured that the collective oscillations are sufficiently damped out and the virial ratio of the bound matter $2K/|W|\simeq 1$. Particles are propagated with a second order leapfrog scheme with adaptive time step (see \citealt{1995ApJ...443L..93H}) $\Delta t$ that varies during each run as $\Delta t\equiv\eta/\sqrt{4\pi G\rho_{\rm max}(t)}$, where $\rho_{\rm max}(t)$ is the time-dependent maximal density and $\eta$ is the so-called Courant-Friedrichs-Lewy condition, that in the simulations discussed here was fixed to 0.3.
\begin{figure*}
    \centering
    \includegraphics[width=0.9\textwidth]{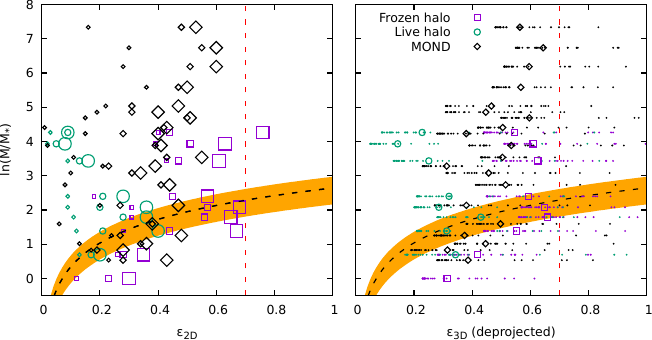}
    \caption{Mass ratio $M/M_*$ as function of the 2D ellipticity $\epsilon_{2D}$ measured on three random projections (indicated with increasingly bigger symbol size) for the end products of models with initially cold ($K_0=0$) Hernquist profiles ($\gamma=1$) with frozen (squares) and live (circles) Hernquist DM halos and in MOND (diamonds) (cfr. rows 7,9 and 22 in Tab. \ref{tab_sim}), and as function of the deprojected ellipticity for different choices of the inclination angle $\theta$ (dots) an their average value (right panel). For comparison, in both panels the dashed line and the orange shaded area mark the \citealt{2014MNRAS.438.1535D} relation, while the vertical red dashed lines indicate the limit ellipticity 0.7.}
    \label{figrgrav}
\end{figure*}
\begin{figure}
    \centering
    \includegraphics[width=\columnwidth]{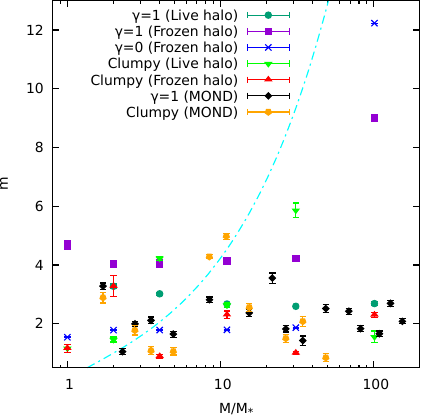}
    \caption{S\'ersic index $m$ as function of the total to baryonic matter ratio for the end products of collapses  with frozen and live DM halos, as well as dynamical to baryonic matter in MOND simulations for the simulations with parameters given in rows (7,9,18,19,22) of Tab. \ref{tab_sim}. The cyan dotted-dashed line marks the relation emerging from the observational data of \citealt{2019A&A...622A..30S}.}
    \label{figrgrav2}
\end{figure}
\begin{figure*}
    \centering
    \includegraphics[width=0.9\textwidth]{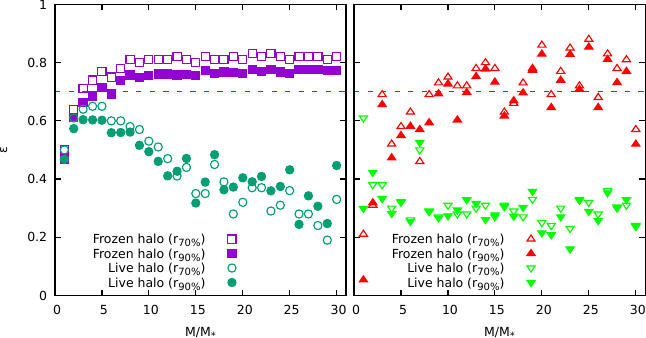}
    \caption{Intrinsic ellipticity $\epsilon$ as function of $M/M_*$ for initially cold spherical systems in live (circles) and frozen (squares) halos (left panel); and for initially cold clumpy systems in live (downward triangles) and frozen (upward triangles) halos (right panel). The empty and filled symbols refer to $\epsilon$ evaluated inside $r_{70\%}$  and $r_{90\%}$, respectively. The simulations' parameters are summarized in rows (14-17) of Tab. \ref{tab_sim}.}
    \label{figrgrav4}
\end{figure*}
\begin{figure*}
    \centering
    \includegraphics[width=\textwidth]{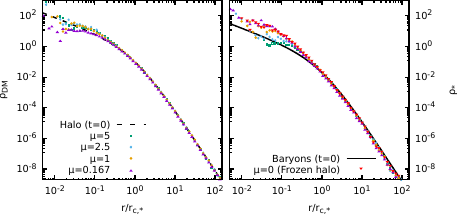}
    \caption{Density profiles at $t=300t_{\rm Dyn}$ for halo (left panel) and baryons (right panel) for $\mu=5$ (squares), 2.5 (circles), 1 (diamonds), 0.167 (triangles) and 0 (downward triangles). The dashed and solid lines mark the initial profiles ($\gamma=1$ in both cases) for DM and baryons, respectively. The simulations' parameters are summarized in row (20) of Tab. \ref{tab_sim}.}
    \label{figrgrav7}
\end{figure*}
\begin{figure*}
    \centering
    \includegraphics[width=0.9\textwidth]{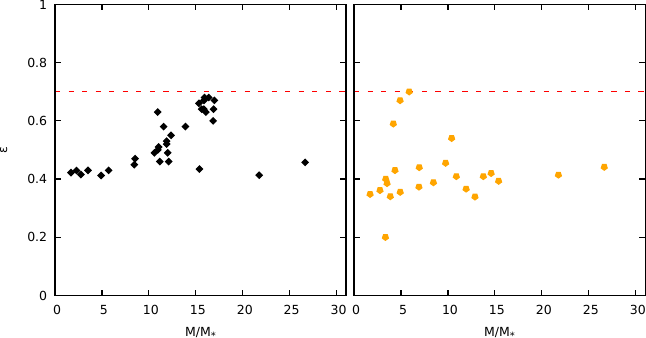}
    \caption{Same as Fig. \ref{figrgrav4} for MOND systems with $\gamma=1$ (left) and clumpy (right) initial conditions, corresponding to the simulation parameters summarized in rows (22-23) of Tab. \ref{tab_sim}.}
    \label{figrgrav9}
\end{figure*}
\begin{figure}
    \centering
    \includegraphics[width=\columnwidth]{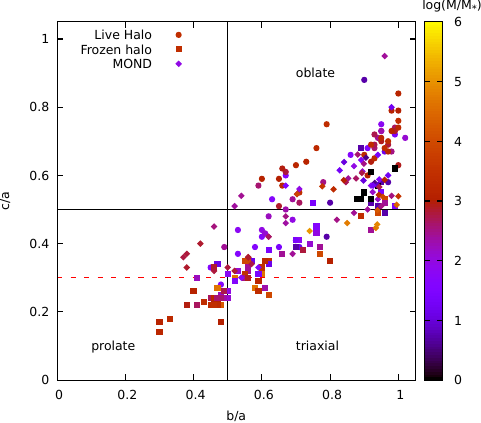}
    \caption{Minimum vs maximum axial ratios for the end products of Newtonian collapses in live (circles) and frozen (squares) halos, and MOND collapses for the different values of the logarithm of the mass ratio $M/M_*$ listed in Col. 4 of Tab. \ref{tab_sim} (colour bar).}
    \label{figrgrav1}
\end{figure}
\begin{figure}
    \centering
    \includegraphics[width=\columnwidth]{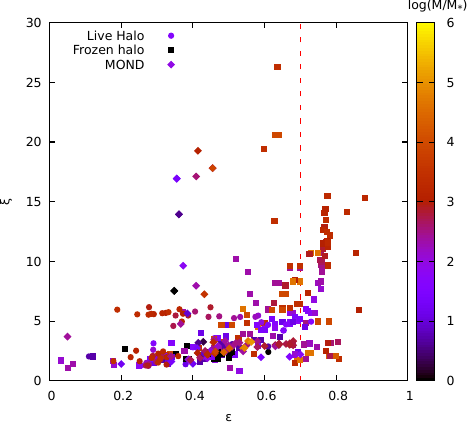}
    \caption{Anisotropy parameter inside $r_{70\%}$ as function of $\epsilon$ for the end products of the simulations. As in Fig. \ref{figrgrav1} the color code maps the (logarithmic) mass ratio $M/M_*$ listed in Col. 4 of Tab. \ref{tab_sim}.}
    \label{figrgrav3}
\end{figure}
\subsection{Structural analysis of the end products}
For the sets of simulations introduced above, projected and intrinsic properties were evaluated in the standard way (see e.g. \citealt{2006MNRAS.370..681N,2013MNRAS.431.3177D} and references therein). Once the end products are translated to the centre of mass frame, the particles undergo three random rotations, the rotated system is then projected in the plane perpendicular to the three putative lines of sight. The 2D ellipticities are obtained as $\epsilon_{2D}=1-\varrho$, where in each of the three projections $\varrho=b_{2D}/a_{2D}$ is the ratio between the minimum and maximum projected semiaxis.\\
\indent To have a measure of the concentration of the model, the angle averaged two dimensional density profiles $\Sigma(R)$ were fitted with the \cite{1968adga.book.....S} law
\begin{equation}\label{sersic}
\Sigma(R)=\Sigma_e e^{-b\left[\left(\frac{R}{R_e}\right)^{1/m}-1\right]}.
\end{equation}
In the equation above $\Sigma_e$ is the projected mass density at effective radius $R_e$, i.e. the radius of the circle containing half of the projected mass. Since the two dimensionless parameters $b$ and $m$ are related by $b\simeq2m-1/3+4/405m$ (see \citealt{1999A&A...352..447C}), Eq.(\ref{sersic}) can be fitted with the single parameter $m$. I recall that, for high values of $m$ the density profile is steep in the central regions and shallow in the outer,
 while low values of $m$ correspond shallow central density profiles with steeper outer slopes.\\
\indent The intrinsic triaxiality of the end products is recovered evaluating the second order tensor\footnote{Note that $I_{ij}$ is {\it not} the inertia tensor, which is given instead by ${\rm Tr}(I_{ij})\delta_{ij}-I_{ij}$.}
\begin{equation}
I_{ij}\equiv m\sum_{k=1}^N r_i^{(k)}r_j^{(k)} 
\end{equation}
for the particles inside the spheres of radius $r_{50}$ , $r_{70}$ and $r_{90}$ containing the 50\%; 70\% and the 90\% of the stellar mass of the system, respectively.\\  
\indent The matrix $I_{ij}$ is diagonalized iteratively, with the requirement that the percentage difference of the largest eigenvalue between two consequent iterations does not exceed $10^{-3}$. On average, for $N\approx 10^5$ the process requires about 10 iterations. Once the three eigenvalues $I_1\geq I_2\geq I_3$ are recovered, a rotation is applied to the system such that the three eigenvectors oriented along the coordinate axes. For of a heterogeneous ellipsoid of semiaxes $a,b$ and $c$, one has $I_1=Aa^2$, $I_2=Ab^2$ and $I_3=Ac^2$, where $A$ is a constant depending on the density profile. The axial ratios are defined by $b/a=\sqrt{I_2/I_1}$ and $c/a=\sqrt{I_3/I_1}$, so that the ellipticities in the principal planes are $\epsilon_1=1-\sqrt{I_2/I_1}$ and $\epsilon_2=1-\sqrt{I_3/I_1}$. Models with $c/a\sim b/a\lesssim 0.5$ are defined prolate while models for which $c/a\sim b/a\gtrsim 0.5$ are defined oblate, while models having $b/a>0.5$ and $c/a<0.5$ result evidently triaxial.\\
\indent For the end products of the MOND simulations, the (bona fide) ENS is recovered with the same procedure of \cite{2023arXiv230708865R} by evaluating the angle averaged (spherical) density profile on a radial grid from which the Newtonian and MOND force fields $\mathbf{g}_N$ and $\mathbf{g}_M$ are recovered. The DM density of the ENS is then obtained as
\begin{equation}\label{rhodmens}
\Tilde{\rho}_{DM}=(4\pi G)^{-1}\nabla\cdot(\mathbf{g}_M-\mathbf{g}_N).
\end{equation}
Note that, Equation (\ref{rhodmens}) above, is valid only for isolated spherical systems, as substituting the source term in Eq. (\ref{MOND}) with the Poisson Equation $\Delta\Phi_N=4\pi G\rho$ and integrating out the divergence term on both sides yields in any other geometry
\begin{equation}
\label{gmgn}
  \mu\left(\frac{||\mathbf{g}_M||}{a_0}\right)\mathbf{g}_M=\mathbf{g}_N+\mathbf{S},
\end{equation}
where $\mathbf{S}\equiv\nabla\times\mathbf{h}(\rho)$ is a density-dependent solenoidal field.\\
\indent The DM content $M_{DM}$ of the ENS is finally obtained by integrating Eq. (\ref{rhodmens}) from 0 up to the radius of the farmost particle so that the total dynamical mass of the model is again $M=M_*+M_{DM}$.
\section{Simulations and results}
\subsection{Projected properties}
Figure \ref{figrgrav} (left panel) shows for three random projections (indicated with differently sized symbols) the 2D ellipticites of the end products of spherical collapses in frozen (squares) and live (circles) halos as well as baryon-only MOND (diamonds) collapses versus the mass ratio on a log scale. In all cases both baryons and DM (when present) start from Hernquist density profiles with the same scale radius $r_s$ and $N=10^5$. The stellar systems produced in collapses within frozen halos show a somewhat increasing trend of $M/M_*$ with $\epsilon_{2D}$ (or vice versa), though not reproducing the linear relation and its associated uncertainty, given in Eq. (\ref{epstoml}) and marked in figure by the dashed line and the (orange) shaded area. Collapses in initially virialized live halos, seem to produce systems with lesser departure from the spherical symmetry at low or rather large (up to $\sim 10^2$) $M/M_*$, with maximum value of $\epsilon_{2D}$ for $M/M_*\sim 3$. Remarkably, for the random projections shown here, at fixed mass ratio $\epsilon_{2D}$ is systematically smaller for the products of the collapses in live halos. MOND collapses show a somewhat intermediate behaviour attaining large values of the projected ellipticity (around $\epsilon_{2D}\sim 0.6$) for dynamical to baryon mass ratios of order $10^2$.\\
\indent For comparison, in the right panel of Fig. \ref{figrgrav} the mass ratio is plotted against the estimated intrinsic minimal ellipticity $\epsilon_{3D}$ recovered inverting Eq. (\ref{epsilonapp}) for the intermediate value of $\epsilon_{2D}$ in the assumption of oblateness for a Gaussian distribution of the assumed projection angle $\theta$ (points). The mean value of $\epsilon_{3D}$ for 30 independent realizations of $\theta$ is marked by the symbols, with the same coding as in the left panel. Again, the dashed curve and the orange shaded area highlight Eq. (\ref{epstoml}). If on one hand the point distributions presents an increasing trend of $M/M_*$ with $\epsilon_{3D}$, on the other, the deprojected ellipticities still fail to reproduce the linear relation observed by \cite{2014MNRAS.438.1535D} and \cite{2023MNRAS.518.2845W}. It is worth noting though, that, some MOND systems and Newtonian systems with a frozen halo are intrinsically prolate (see discussion below), thus contradicting the oblateness assumption in applying Eq. (\ref{epsilonapp}).\\
\indent Final states attained starting by different sets of Newtonian initial conditions (i.e. clumpy, $\gamma=0$ flat cored initial baryon profiles, not shown here) do not present radically different projected ellipticity trends. Figure \ref{figrgrav2} presents the Sersi\'c index $m$ as a function of $M/M_*$. Most models, independently on the specific value of the mass ratio have indexes in the range $1.1\lesssim m\lesssim 7$, compatible with the values for the observed elliptical galaxies (e.g. see \citealt{2017ApJ...841...32Z,2019A&A...622A..30S,2019lgei.confE..64S} and references therein). Collapses of initially spherical distributions in frozen halos, produce remarkably large Sersi\'c indexes at high $M/M_*$, with values up to $\sim 9$ and 12.3 for initially cuspy $\gamma=1$ or flat cored $\gamma=0$ baryon profiles, respectively. For $M/M_*\lesssim 40$ the $\gamma=1$ initial condition in a frozen halo produce final values of $m\sim 4$, corresponding to a \cite{1948AnAp...11..247D} profile.\\
\indent The S\'ersic indexes of MOND models starting from both spherical (diamonds) and clumpy (pentagons) initial conditions do not show a clear trend with the ratio of dynamical to stellar mass, with initially spherical systems yielding a narrower span of values around $m\sim 2$. Notably, MOND clumpy initial conditions could produce extremely centrally shallow final projected profiles with $m$ down to $\approx 0.89$ for $M/M_*\approx 50$ (corresponding to $\kappa\sim 10$). \cite{2019A&A...622A..30S} found for their sample of spheroids that $m\propto M_*^{0.46}$ and $M_{DM}\propto M_*^{1.7}$, that would correspond roughly to $m\propto (M/M_*-1)^{0.66}$, indicated in Figure by the dotted-dashed line. With the notable exception of the spherical collapses in frozen halos whose end products exhibit and increasing trend of $m$ with $M/M_*$, the other sets of simulations fall outside the trend of the S\'ersic index with $M/M_*$ when the latter exceeds $\approx 30$. Notably, $M/M_*\approx 30$ is also the maximum value of the mass ratio explored in the samples use d by \cite{2019A&A...622A..30S}. 
\subsection{Intrinsic properties}
As in the $N-$body simulations discussed here the ratio of the total dynamical mass to stellar mass is essentially the control parameter, it is convenient to show the intrinsic minimum ellipticity $\epsilon$ as function of $M/M_*$. In Fig. \ref{figrgrav4} $\epsilon$ is plotted against $M/M_*$ for Newtonian simulations with spherical ($\gamma=1$, left panel) and clumpy (right panel) initial conditions for the baryons. In all cases the DM halo is, at least initially, simulated with a $\gamma=1$ model.\\
\indent Initially spherical systems collapsing in a frozen halo with central $\rho_{DM}\propto1/r$ (indicated in figure by empty and filled squares) relax to flatter end states for increasing $M/M_*$, eventually tending to $\epsilon\approx 0.8$. For mass ratios $M/M_*\gtrsim 5$ such end states are significantly more flattened than a E7 galaxy (indicated by the dashed line at $\epsilon=0.7$) both at the Lagrangian radii enclosing 70\% (empty symbols) and 90\% (filled symbols) of $M_*$. When the halo is live (i.e., modelled using particles, circles in figure) the picture is rather different and the ellipticity shows a non monotonic trend with the mass ratio $M/M_*$. Curiously, $\epsilon$ starts to decrease with at around $M/M_*\approx 5$, value for which $\epsilon(M/M_*)$ settles to a somewhat constant value for the frozen halo simulations. Spherical collapses in cored halos (i.e. with $\gamma=0$, not shown here) have essentially the same behaviour as their counterparts with a cusp, though for the cases where said halo is frozen, the limit value of $\epsilon$ at increasing $M/M_*$ is at around 0.7.\\
\indent Clumpy initial conditions in spherical halos yield end products that have a qualitative trend of $\epsilon$ with $M/M_*$ as their spherical counterparts when the DM halo is frozen (though with a larger scatter of $\epsilon$ at bigger mass ratios). Vice versa, a live halo almost always induces mildly flattened end states, around $\epsilon\approx 0.3$ for $M/M_*\gtrsim 6$ while a somewhat non-monotonic trend is evident a low mass ratios.\\
\begin{figure*}
    \centering
    \includegraphics[width=0.9\textwidth]{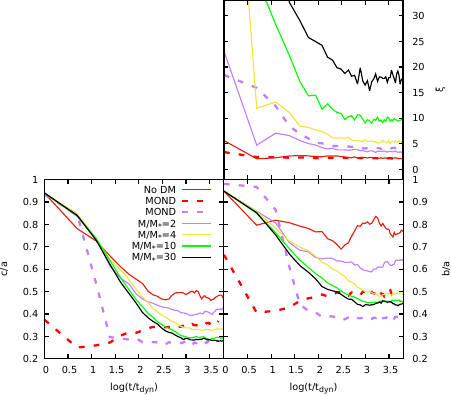}
    \caption{Evolution of the anisotropy parameter $\xi$ (top right panel) and axial ratios $c/a$ (bottom left) and $c/a$ (bottom right) for different Newtonian models in frozen halos (solid lines, row 7 in Tab. \ref{tab_sim}) and MOND models (dashed lines)}
    \label{figrgrav10}
\end{figure*}
\indent The final DM halos' ellipticities $\epsilon_{DM}$ have been evaluated for all live halo simulations. In general, for $N_*=N_{DM}=5\times 10^4$ one has $0.93\lesssim \epsilon_{DM}\lesssim 0.98$ implying that the collapse of the baryon distribution does not alter significantly the sphericity of the virialized halo, while its central regions become significantly more ''cored" at later times when $M/M_*\lesssim 6$. This could be in principle a numerical collisionality artifact induced by the simulation resolution (i.e. number of particles used for DM and stars and particle mass ratio $\mu=m_{DM}/m_*$). For example, Figure \ref{figrgrav7} shows the final DM (left panel) and stellar (right panel) density profiles for different realizations of a model with initial $\gamma=1$ halo and stars density profiles with $M/M_*=5$ and various choices of particle resolution ranging from $\mu=5$ to 0.167 (i.e. the individual mass of DM particles spans from values larger than that of particles representing the stellar component to significantly lower). In both sets of density profiles the curves differ significantly from one another at radii smaller than $r/r_{c,*}\sim 10^{-2}$. In particular, the halo density $\rho_{DM}$ becomes increasingly cored for higher DM resolutions\footnote{In principle, the DM resolution-relate issues could be probed using the effective multi-component models introduced by \citealt{2021MNRAS.503.4221N}.} (i.e. larger $N_{DM}$ and and smaller $\mu$ at fixed $M/M_{*}$). Surprisingly, in that case, the associated $\rho_*$ is strikingly similar to that obtained for the frozen halo simulation (downward and upward triangles in the right panel). A similar behaviour has been verified for lower values of $M/M_*$ and clumpy initial conditions for the initial stellar distribution.\\
\indent Figure \ref{figrgrav9} shows $\epsilon$ as function of $M/M_*$ in MOND simulations (corresponding to rows 22 and 23 in Tab. \ref{tab_sim}) where the effective total  dynamical mass $M$ is obtained as discussed in Sect. 2.3. Spherical collapses starting from cuspy initial conditions with $\gamma=1$ and various values of the parameter $\kappa$ always produce relaxed end states with $0.4\lesssim\epsilon\lesssim 0.7$. Between $M/M_*=1$ and $\approx 14$, $\epsilon$ has a markedly increasing trend, though nowhere linear as claimed in \cite{2014MNRAS.438.1535D} and \cite{2023MNRAS.518.2845W} and observed in the MOND $N-$body simulations by \cite{2023arXiv230708865R}. The latter simulations however, where limited to $\kappa=1$ and 100 cases only, but with different values of the initial virial ratio $2K/|W|$. Clumpy systems (as for their Newtonian counterparts) have a rather flat trend of $\epsilon$ with a few outliers for values of the mass ratio smaller than $M/M_*\approx 5$, corresponding to simulations with an initial value of $\kappa\approx 500$. Curiously, for spherical initial conditions, the minimal ellipticity is obtained in correspondence of the trend inversion.\\
\indent Newtonian and MOND systems starting from similar initial conditions have qualitatively different intrinsic triaxiality as summarized in Fig. \ref{figrgrav1} where the final ratio of the minimum to maximum semiaxis $c/a$ is plotted against the ratio of the intermediate to maximum semiaxis $b/a$. In general, as observed also in \citealt{2007ApJ...660..256N,2011MNRAS.414.3298N}, single component Newtonian collapses mostly produce marginally oblate systems while the end products of MOND collapses are in general slightly prolate or triaxial. Here we observe that when a (frozen) DM halo is added (squares in figure), Newtonian end states often become markedly prolate, in particular for $M/M_*> 20$ (as colour coded in Fig. \ref{figrgrav1}), in particular when $c/a$ falls below 0.3 (as indicated by the red dashed line), corresponding to ellipticities larger than the threshold value of $\epsilon=0.7$ for an E7 galaxy. Newtonian collapses in live halos (cfr. circles) mostly produce oblate (at large $M/M_*$) or mildly triaxial systems (for low values of $M/M_*$). The MOND simulations performed here typically produce oblate end states for effective $M/M_*\lesssim 6$ and markedly prolate end states for $6\lesssim M/M_*\lesssim 35$. Larger values of the effective mass ratio (associated to values of $\kappa<10$ in the initial condition) generally produce strongly triaxial systems, without any evident trend with the cored, cuspy or clumpy nature of the initial mass distribution.\\
\indent Independently on the specific gravity model at hand (Newtonian or MOND), as a general trend the simulations performed in this work yield more and more radially anisotropic states for larger values of $\epsilon$, up to $\xi\sim 26.5$ for Newtonian models collapsing in frozen halos. Typically, increasing values of $M/M_*$ reflect in larger final $\xi$. In Figure \ref{figrgrav3} (cfr. also Fig. 7 in \citealt{2023arXiv230708865R}) the anisotropy parameter $\xi$ is shown as function of $\epsilon$ for various choices of $M/M_*$ indicated by the color map. MOND models stand out as being able to attain rather large values of $\xi$ for intermediate values of $\epsilon$ (around $0.45$) and small values of their effective dynamical to stellar mass ratio $M/M_*$. 
As shown by the time evolution of $\xi$ (top panel) and the axial ratios (bottom panels) of Fig. \ref{figrgrav10}, for $\gamma=1$ systems, MOND models (red and purple dashed lines) reaching comparable values of the anisotropy index to that of Newtonian systems with a frozen halo\footnote{Newtonian models with a live halo could in principle exchange orbital anisotropy between the two components via spurious (numerical) collisional processes, in particular in the central regions where the resolution is, by construction, scarce.} (line 7 in Tab. \ref{tab_sim}) become considerably more flattened in lesser time in units of their dynamical time. Note that, for a given stellar mass $M_*$, MOND systems have significantly different dynamical time scales than comparable Newtonian models with a DM halo (cfr. Eq. \ref{tdyn}), with MONDian collective time-scales being typically longer (\citealt{2007ApJ...660..256N}). Note also that, MOND systems attaining similar values of $\xi$ to the Newtonian models with $M/M_*=2$ and 1 have slightly larger ratios of dynamical to stellar mass. ($\approx 4$ and $1.7$, respectively). One has also to bear in mind that, when converted in physical units, the time scales of Fig. \ref{figrgrav10} may differe significantly from one another. For example, using the halo to stellar mass relation of \cite{2019A&A...622A..30S} (see also \citealt{2017ApJ...839..121T}), with a scale ratio of 100 kpc, the dynamical time for Newtonian (baryons plus DM) systems would range from $\approx 1.3$ Gyrs down to $\approx0.06$ Gyrs for mass ratios in the range $2\lesssim M/M_*\lesssim 30$, spanning an interval of luminous masses between $10^{10}$ and $10^{12}~M_{\odot}$.\\ 
\indent As expected, in both paradigms of gravity explored here, initially cold spherical models ($2K/|W|\ll 1$), become rapidly (i.e. below $1t_{Dyn}$) radially anisotropic and undergo a process akin to the ROI for unstable equilibrium systems and once relaxed appear triaxial or flattened. In the Newtonian cases, the bigger the mass ratio $M/M_*$ the higher the initial (and final) value of $\xi$ (cfr. top panel in Fig. \ref{figrgrav10}). This leads to conjecture that a (almost) spherical pre-formed DM halo induces strongly radially unstable initial states for the stellar/baryonic component, at least within an interval of $M/M_*$. For fixed mass ratios and halo density profile, the end products of cold clumpy initial conditions are systematically less anisotropic than their initially spherical counterparts. In MOND collapses, on the contrary, clumpy initial conditions tend to produce systems with larger values of $\xi$ than those starting spherical, with comparable ''phantom" DM evaluated from Eq. (\ref{rhodmens}), reaching similar final values of $\epsilon$.
\section{Summary and discussions}
Aiming at shedding some light on the origin the ellipticity - dynamical mass relation I have performed $N-$body simulations of dissipationless collapse in both Newtonian gravity with dark matter and MOND. In general, the products of most numerical experiments, in agreement with previous studies with analogous initial conditions, have values of the S\'ersic index, radial anisotropy and density profiles in ranges comparable to those of the observed elliptical galaxies.\\
\indent The Newtonian and MOND simulations here discussed both show that, at least for values of the mass ratio $M/M_*$ (intrinsic or effective in the case of MOND) between 5 and 6, a certain increasing trend with the ellipticity of the spheroid is produced by disiipationless collapse alone. However, the linear proportionality (cfr. Eq. \ref{epstoml}) discussed by \cite{2014MNRAS.438.1535D} and later by \cite{2023MNRAS.518.2845W} could not be recovered in any of the two paradigms of gravity considered here, neither for the intrinsic ellipticities nor their deprojected values estimated with Eq. (\ref{epsilonapp}).\\
\indent In particular, in Newtonian simulations with a frozen DM halo the values of $\epsilon$ at the Lagrangian radii enclosing the 90\% of the stellar mass ''saturate" at around 0.75 for $M/M_*$ larger than 6. Of course, this is most likely a numerical effect due to the rather unphysical fixed halo assumption. For the Newtonian simulations where a live DM halo was considered, an unequivocal non-monotonic relation between $\epsilon$ and $M/M_*$ (and hence $M/L$ if a constant $L/M_*$ is assumed) is observed. In particular, a tendency to produce less flattened end states when $M/M_*$ exceeds $\approx 20$ is observed for spherical initial conditions. It is therefore tantalizing to infer that at even larger mass ratios the end products of Newtonian collapses should essentially be almost spherical, as one would expect for the case of ultrafaint dwarf galaxies where $M/M_*$ may exceed $10^2$ (e.g. \citealt{2021A&A...651A..80Z}). One should however be aware that for a given mass ratio, the end products of live halo simulations could be influenced by the resolution. Indeed, for spherical collapses for increasing resolution (i.e., decreasing values of $\mu$) the final models tend to depart more from the spherical symmetry. MOND models interpreted in the context of a DM scenario (i.e. when the halo of the associated ENS is accounted) also present an increasing trend of $\epsilon$ with $M/M_*$ below $M/M_*\approx 10$, in partial contrast with the simulations of \cite{2023arXiv230708865R}, limited to a single value of $\kappa=GM_*/a_0^2r_s^2$, yielding a monotonic and quasi-linear trend between the quantities.\\
\indent Prompted by recent results on the relation between stellar and halo masses and the S\'ersic index of the stellar component (see \citealt{2019A&A...622A..30S}), $m$ was evaluated for the Newtonian and MOND simulation outputs. Remarkably, MOND collapses (both clumpy and spherical) are found to be able to produce systems with $m$ order unity (corresponding to an extremely shallow core) without invoking any dissipative mechanism, at variance with the Newtonian simulations of \cite{2015ApJ...805L..16N}, where clumpy initial conditions where used with the power-spectrum index $n$ as control parameter. Clumpy initial conditions in Newtonian simulations with frozen halos also can also attain low values of $m$, however in such cases this is likely a numerical artifact. Independently on the specific nature of the initial baryon distribution, live DM halos flatten down to $\epsilon\sim 0.06$ (retaining a oblate 3D structure) and have their central cusp significantly lowered for clumpy initial conditions on the stellar mass, thus qualitatively confirming what noted in the two component Newtonian simulations of \cite{2011MNRAS.416.1118C}, see also \cite{2023MNRAS.526.1428P}. The increasing trend of $m$ with $M/M_*$ deducible from \cite{2019A&A...622A..30S}, is observed only below $M/M_*\sim 30$. However, systems with larger dynamical to luminous matter where not in the samples there analyzed.\\
\indent To summarize, the above listed results point toward the fact that in a certain span of halo masses in units of the visible matter total mass $M_*$ the increasing trend between $M/M_*$ and the flattening $\epsilon$ could be a consequence of the dissipationless collapse. The fact that a unequivocal linear relation is not recovered, can be likely ascribed to the uncertainty on the measurements or the deprojection procedure for $\epsilon$, and/or to the strong assumptions made on the mass to light ratio of the visible matter. Moreover, the non-monotonicity in the trend between $\epsilon$ and $M/M_*$ in some simulation set-ups is not at all in contrast with the observed ellipticals. The qualitative interpretation that emerges is that, the deeper the (total) potential well during the collapses, the stronger is the radial orbit instability and hence the departure from the spherical symmetry of the final state. For extremely large $M/M_*$, the baryonic matter behaves essentially as tracer particles within the halo potential. If the latter has a rather spherical shape, the stellar component hardly departs from it, as no ROI takes place.\\
\indent Most importantly, one has to bear in mind that Eq. (\ref{epstoml}) was defined for an ensemble explicitly excluding systems, such as ultrafaint dwarfs, that by definition fall outside an increasing trend of $M/M_*$ with $\epsilon$. From the point of view of modified Newtonian dynamics (MOND), evaluating the dark mass content of the ENS of the end states of MONDian simulations reveals that a relation akin to Eq. (\ref{epstoml}) could be also supported in a modified dynamics scenario. As a by product of this work, it appears that in MOND sufficiently clumped initial conditions can yield 
 flattened and centrally shallow spheroids even in absence of dissipation, as opposed to Newtonian collapses with DM requiring some form of dissipation.\\
 \indent A natural follow-up of this study, in the context of Newtonian gravity, would focus necessarily on the effects of an initially non-virialized halo as well as a clumpy DM distributions, not considered here for reasons of simplicity. For what concerns MOND (and possibly other alternative theories without DM), the interplay between the clumpyness of the initial (baryonic) density and the initial acceleration regime (i.e. how below or above $a_0$) should also be explored.
\begin{acknowledgements}
I would like to express gratitude to Federico Re, Carlo Nipoti and Stefano Zibetti for the useful discussions at an early stage of this work. The Referee is also acknowledged for his/her suggestions that helped improving the presentation of the results. I am also wishing to acknowledge funding by ‘‘Fondazione Cassa di Risparmio di Firenze" under the project {\it HIPERCRHEL} for the use of high performance computing resources at the University of Firenze.
\end{acknowledgements}
   \bibliographystyle{aa} 
   \bibliography{biblio} 
\end{document}